# Do cell culturing influence the radiosensitizing effect of gold nanoparticles part 2: scrutinizing the methodology producing recent evidence

Hans Rabus[1], Oswald Msosa Mkanda[1,2,3]

[1] Physikalisch-Technische Bundesanstalt, Berlin, Germany
[2] University Dar es Salaam, Dar es Salaam, Tanzania
[3] Mzuzu University, Mzuzu, Malawi

E-mail: hans.rabus@ptb.de

**Abstract**

When irradiation is performed in the presence of dose modifying agents such as gold nanoparticles (AuNPs), a different shape of a cells in suspension or adherent to container walls may result in different dose to the nucleus and different probability of cell survival. In a recent simulation study based on voxelized cell models, differences of up to a factor of 2 were found between the predicted survival of floating and adherent cells or cells oriented differently with respect to the incident beam. In the first part of the paper, it was shown that there are inconsistencies in the reported data and a bias in the simulation setup. Indeed, only cells near the surface of a volume were considered, so the simulation may not be representative of the experiments it aims to reproduce. The present work aims to quantify the biases introduced by the simulation setup and the use of voxelized geometry in conjunction with the local effect model for cell survival. The results show that simulated irradiation of a cell near the surface with an incident beam matched to the cell dimensions results in dose values that are by a factor of about 50 lower than the dose to cells deeper in the medium when irradiated with a $^{60}$Co spectrum and lateral beam dimensions in the centimeter range. Furthermore, the number of ionizing photon interactions in gold nanoparticles in a cell near the surface is lower by a factor of about 2 than for cells at 5 mm and 1 cm depth. Using the average dose in voxels of dimensions in the order of 0.2 μm for the assessment of cell survival according to the local effect model (LEM) leads to an underestimation of the number of lesions from a single AuNP undergoing an ionizing interaction by roughly two order of magnitude and thus to an overestimation of cell survival. These results emphasize that a realistic approximation of the secondary radiation field and the dose distribution in the macroscopic sample must be ensured by the simulation setup in order to avoid strong bias in the results. In addition, the effect of cell geometry on the predicted survival rate was examined for approximate cell geometries for 100 kV x-ray irradiation, for which the probability of photon interaction in gold nanoparticles is by more than two orders of magnitude higher than for $^{60}$Co irradiation. The results show that the effects are negligible for 5 nm nanoparticles at the concentration of considered in preceding work. For 50 nm nanoparticles and thus a thousand times higher mass fraction of gold, significant reduction in cell survival is found, with a clear additional reduction predicted by the LEM as compared to the prediction based on mean dose to the nucleus.

## 1. Introduction

The geometry of biological cells irradiated in vitro may differ when they are floating in the surrounding medium or adherent to a surface such as a biofoil. For floating cells, surface tension favors a more spherical shape whereas for adherent cell, surface tension is minimized when the cell is flattened out. When cells are irradiated in the presence of dose-modifying agents such as nanoparticles (NPs) of a material with high atomic number Z, the dose distribution in the cell becomes non-uniform. As was shown by Sung *et al* (2017), the resulting cell survival probability depend on the spatial distribution of the NPs in and around the cell (Sung et al., 2017). Therefore, a different shape of the cell may have an impact on the cell survival probability when NPs are present.

A recent simulation study by (Antunes et al., 2025a) reported the predicted survival of cells containing gold NPs (AuNPs) to strongly depend on cell morphology and ón the direction of irradiation (for adherent cells). The simulations were based on detailed voxelized models of the cells derived from confocal microscopy imaging (Antunes et al., 2024). The considered irradiations were in a $^{60}$Co irradiation facility and in the Bragg-peak region of a proton beam of 14 MeV initial energy.

In the first paper (Rabus and Mkanda, 2025), inconsistencies in the data reported by (Antunes et al., 2025a) for the $^{60}$Co irradiation were pointed out that may in part be explained by bias in the simulation. For instance, the simulation setup corresponds to the case of cells at a surface where dose-buildup results in a steep dose gradient that can explain different survival for different cell orientation or shape (Rabus and Mkanda, 2025). Another inconsistency was that the datapoints for survival predicted from the simulation results were following a straight line in the conventional semi-logarithmic plot instead of a parabola as would be expected according to the linear-quadratic (LQ) model of cell survival. Assuming that (Antunes et al., 2025a) neglected the quadratic term of the LQ model, corrected survival curves were determined in (Rabus and Mkanda, 2025). These corrected curves failed to reproduce the experimental data.

In this part of the paper, the methodology used by (Antunes et al., 2025a) is examined with the aim of quantifying the bias to the simulation results from the irradiation setup. In the simulations, the radiation source was placed next to the cells and its dimensions were matched to the cell cross-section.

There are two important points to note with this simulation approach: First, when the lateral extent of the radiation field is limited to the cell dimensions, a large fraction of the contribution of backscattered photons that are produced downstream is suppressed. In addition, there is no lateral charged particle equilibrium, which leads to biased results (Rabus et al., 2021).

The second point is that this geometry setup is equivalent to considering only cells near the surface of the MWP or, if the wells were not filled to the top, from the surface of the medium containing the cells in a well. However, the radiation field impinging on the surface of the plate is not the same as that impinging on cells in the volume, as secondary particles produced upstream are omitted so that the secondary particle component of the radiation field is much smaller than at greater depths.

Apart from these issues with the simulation setup, the predictions for cell survival probability according to the LEM in (Antunes et al., 2025a) are based on dose values in voxel of dimensions in the range of 0.21 μm to 0.6 μm. This means that they are larger than the radial range around nanoparticles in which order-of-magnitude variations of the local dose occur (McMahon et al., 2011a, 2011b). This implies that there may be an underestimation of the LEM-specific contribution to the expected number of lesions that determines the cell survival probability. This contribution is related to the average square of the dose around a nanoparticle. Its accurate determination may require using voxel sizes in the few nanometer range (Lin et al., 2015; Velten and Tomé, 2024, 2023) or analytical approaches (Melo-Bernal et al., 2021; Rabus, 2024; Rabus and Thomas, 2025). Determining the bias from using a coarse voxel grid is therefore another aim of this work.

## 2. Materials and Methods

### 2.1 Indicent fluence for the $^{60}$Co cell irradiation

(Antunes et al., 2025a) used a two-step simulation approach. In the first step, the particles hitting the rectangular front side of a multi-well plate (MWP) were scored using an irradiation geometry mimicking that of the $^{60}$Co irradiator facility. The in-plane coordinates in the resulting phase-space file (PSF) were then multiplied by factors (one per dimension) such that the rectangle of the MWP surface was transformed into a smaller rectangle having a cross section "matching that of the cell model" (Antunes et al., 2025a). The resulting transformed PSF was used in the second simulation as the radiation source by random sampling from its entries.

The outcome of this second simulation is the dose per event sampled from this PSF. To obtain the dose per decay of the $^{60}$Co source, (Antunes et al., 2025a) multiplied the dose per event by the number of incident particles calculated with Eq. (1).

$$N = \frac{S_{cell}}{S_{plate}} \times t \times A \quad (1)$$

Here, $S_{cell}$ is the cross-section of the cell, $S_{plate}$ is the surface area of the MWP, $t$ is the irradiation time, and $A$ is the activity of the $^{60}$Co source. The first factor on the right-hand side of Eq. (1) compensates for the increase of particle fluence implied in the transformation of the PSF. The irradiation time is related to the dose rate $\dot{D}$ at the $^{60}$Co irradiation facility and the dose $D$ to be applied in the experiments by $t = D/\dot{D}$.

There are two important points to note with the approach taken by (Antunes et al., 2025a). The first is that the number of photons impinging on the cell is estimated from a dose applied in the experiments and then used to calculate the dose in the simulation. For an irradiation in the absence of AuNPs, this dose calculated from the simulation results should be the same as the corresponding dose in the experiments.

The second point is that Eq. (1) implicitly assumes that each decay results in particles hitting the area $S_{plate}$. The number of decays resulting in particles impinging on the well-plate surface could be easily determined by analyzing the PSF from the first simulation. However, the information on the number of events contained in the PSF is not provided by (Antunes et al., 2025a), and it must be assumed that it was not determined.

Therefore, the bias resulting from the assumption that each decay of the $^{60}$Co source leads to a particle hitting the MWP is estimated here by three different approaches. In the first approach, the photon source is approximated as a point source emitting only the two main $^{60}$Co gamma lines. In this case, the probability of a photon hitting the MWP upon a decay is twice as large as the proportion of the full solid angle subtended by the area $S_{plate}$. The factor 2 takes into account that two gamma photons are emitted per decay.

The second approach is a refinement of the first one and considers that the flux of photons passing through the area $S_{plate}$ has two components: First, $^{60}$Co gamma photons that are emitted into the solid angle subtended by the area $S_{plate}$ and do not interact on their way to the MWP. Second, secondary photons, which are produced by Compton scattering of photons or bremsstrahlung interaction of secondary electrons. These interactions take place in the $^{60}$Co source, the air between the source and the MWP, and in the walls of the irradiator chamber.

Since the emission of the two gamma photons in a $^{60}$Co decay is isotropic, the number $n_{pd}$ of photons hitting the MWP per decay in the $^{60}$Co source can be estimated using Eq. (2).

$$n_{pd} = 2 \times \frac{S_{plate}}{4\pi(d_a + d_s)^2} \times e^{-\mu_s d_s - \mu_a d_a} \times (1 + G_{sec}) \quad (2)$$

Here, $d_s$ and $d_a$ denote the average path length that non-interacting $^{60}$Co gamma photons travel in the source and in the air, respectively, before they reach the front of the MWP. $\mu_s$ and $\mu_a$ are the linear attenuation coefficients of the two materials for photons of the corresponding energies. The factor 2 in front considers that two gamma photons are emitted per decay, and the term $G_{sec}$ denotes the ratio of the number of secondary photons to primary photons hitting the MWP. The term $G_{sec}$ is estimated from the photon energy spectrum of the $^{60}$Co calibration facility at the German National Metrology Institute (PTB), shown in Fig. 2.

The third approach uses the information of the dose rate measured at the $^{60}$Co facility used by (Antunes et al., 2025a). According to (Rabus et al., 2021), the dose rate $\dot{D}_w$ is related to the incident fluence rate $\dot{\Phi}_p$ by Eq. (5):

$$\dot{D}_w = \dot{\Phi}_p \int E \times \left(\frac{\mu_{en}}{\rho}\right)_w \Phi_{rel}(E) dE \quad (3)$$

Here, $(\mu_{en}/\rho)_w$ is the mass energy absorption coefficient of water, $\Phi_{rel}(E)$ is the relative spectral fluence (normalized to unity), and $E$ is the photon energy.

The integral was determined for the photon energy spectrum shown in Fig. 2 using the generalized spline functions from (Rabus et al., 2019) for $(\mu_{en}/\rho)_w$ and for the case of a source emitting only the main $^{60}$Co gamma lines. The proportion of decays in which a photon hits the surface of the MWP is then determined by Eq. (4):

$$n_{pd} = \frac{\dot{\Phi}_p \times S_{plate}}{A} \quad (4)$$

## 2.2 Simulations of the secondary particle field

The particle-energy spectra used in the second simulation could have been extracted from the PSF, but were not shown in the paper of (Antunes et al., 2025a). For lack of possibility to determine the bias from the different radiation fields at the cells, existing data of photon and electron fields are used for obtaining indicative estimates. These data were generated by simulations of a pencil beam of photons with the energy spectrum of the PTB $^{60}$Co calibration facility using the PENELOPE (Penetration and energy loss of photons and electrons) code version 2018 (Salvat, 2019). A customized version of the "Pencyl" code provided with the PENELOPE package was used as the master program.

In these simulations, a pencil photon beam was impinging on a cylindrical water phantom of 15 cm radius and 2 cm depth. The spectral photon and electron fluence were scored in concentric cylinders centered at a depth of 50.5 μm, 5 mm, and 1 cm below the surface. The length of the cylinders along the beam direction was 100 μm, and their radii were 10 μm, 25 μm, 50 μm, 100 μm, 500 μm, 1 mm, 5 mm, and 5 cm. The first and last energy of the fluence histograms were 5 keV and 1345 keV, respectively, and the width of the energy bins was 1 keV.

The simulations were performed with $10^8$ primary particles using condensed-history mode with parameters C1 and C2, i.e. the maximum fractional energy loss and momentum transfer, both set to 0.05. The photon production thresholds (WCR) and the absorption energies of photons were set to 100 eV for the entire geometry. For electrons and positrons, the absorption energies and production thresholds (WCC) were set at 100 eV in the region between the front surface and a plane 100 μm downstream from the back of the cylinders and at 10 keV in the remaining part of the geometry.

## 2.3 Number of hits and ionizing interaction in AuNPs

For the photon fluence spectra obtained by the PENELOPE simulations, the ratio of the expected number $n_i$ of ionizing photon interactions in an AuNP to photon fluence $\Phi_p$ was determined with Eq. (5):

$$\frac{n_i}{\Phi_p} = \int \left(\frac{\mu_i}{\rho}\right)_g \Phi_{rel}(E) dE \times \rho_g V_{np} \quad (5)$$

Here, $(\mu_i/\rho)_g$ is the mass attenuation coefficient for ionizing interaction in gold, $\Phi_p$ is the total photon fluence at the AuNP, $\Phi_{rel}(E)$ is the relative spectral fluence (normalized to unity), and $\rho_g$ and $V_{np}$ are the density of gold and volume of the AuNP, respectively.

The numerical evaluation of Eq. (5) applies a development of the approach by (Rabus et al., 2019), in which a generalized spline interpolation of the mass attenuation coefficient is employed as weighting factor for the photon fluence. Calculating the weighted sum over the bins is then used for calculation of the integral given in Eq. (5). The development is that a generalized spline function for the attenuation coefficient of coherent scattering in gold was determined and subtracted from the total interaction coefficient to only determine interactions leading to an ionization in the AuNP.

With the results from Eq. (5), the expected number $n_{i,1}$ of ionizing interactions in an AuNP per photon impinging on it can be obtained by Eq. (6):

$$n_{i,1} = \frac{n_i}{\Phi_p} \times \frac{1}{A_{np}} \quad (6)$$

In the simulations of Antunes *et al* a number $N_0$ of photons was emitted from the source of area $A_s$. The corresponding fluence for an extended source is $N_0/A_s$. The expectation for the number of photons hitting one out of $N_{np}$ AuNPs is then given by Eq. (7):

$$N_p = N_{np} \times A_{np} \times \frac{N_0}{A_s} \times \frac{\Phi_p}{\Phi_0} \quad (7)$$

Here, $\Phi_p/\Phi_0$ is the ratio of the photon fluence in the cell to the incident fluence. Replacing the photon fluence $\Phi_p$ by the electron fluence $\Phi_e$ in Eq. (7) gives the corresponding expected number $N_e$ of electrons hitting an AuNP in the simulation. Eq. (8) gives the corresponding expression for the expected number of ionizing photon interactions in AuNPs in the simulations:

$$N_i = N_p \times n_{i,1} = \frac{n_i}{\Phi_p} \times N_{np} \times \frac{N_0}{A_s} \times \frac{\Phi_p}{\Phi_0} \quad (8)$$

## 2.4 Bias in the LEM from the voxelized geometry

(Antunes et al., 2025a) used the local effect model (LEM) with the linear-quadratic (LQ) dose dependence of lesion induction to determine cell survival probabilities from the simulations with AuNPs. The resulting cell survial probability is given by Eq. (9):

$$S_{\text{LEM}} = e^{-\alpha \bar{D} - \beta \overline{D^2}} \quad (9)$$

In Eq. (9), $\bar{D}$ and $\overline{D^2}$ are the mean dose and the mean square of the dose in the cell nucleus, respectively, defined by Eq. (10) with $D(\vec{r})$ being the local dose at point $\vec{r}$ and $V_n$ the volume of the cell nucleus:

$$\bar{D} = \frac{1}{V_n}\int_{V_n} D(\vec{r})\, dV_n \; ; \; \overline{D^2} = \frac{1}{V_n}\int_{V_n} D^2(\vec{r})\, dV_n \quad (10)$$

In the work of (Antunes et al., 2025a), a voxized geometry of the cell nucleus was used and $\bar{D}$ and $\overline{D^2}$ were determined according to Eq. (11):

$$\bar{D}_v = \frac{1}{n}\sum_{i=1}^{n} D_i \; ; \; \overline{D_v^2} = \frac{1}{n}\sum_{i=1}^{n} D_i^2 \; ; \; D_i = N d_i \quad (11)$$

In Eq. (11), $D_i$ and $d_i$ denote the dose in the $i$ and the corresponding dose per event in the simulation, respectively. $N$ is the number of incident particles, which was determined in (Antunes et al., 2025a) by Eq. (1) but should have been corrected by the factor $n_{pd}$ given in Eq. (2).

Since $D_i$ is the mean dose in the $i$-th voxel, it can be written as follows where $V_v$ is the volume of a voxel and $V_i$ is the region in space occupied by the $i$-th voxel:

$$D_i = \frac{1}{V_v}\int_{V_i} D(\vec{r})\, dV_i \quad (12)$$

In addition, $V_n = n V_v$ so that the first terms in Eqs. (10) and (11) are identical. On the other hand, if the second term of Eqs. (10) is written as a sum over the voxels, Eq. (13) is obtained, which generally differs from the second identity in Eq. (11)

$$\overline{D^2} = \frac{1}{n}\sum_{i=1}^{n}\overline{D_i^2} \; ; \quad \overline{D_i^2} = \frac{1}{V_v}\int_{V_i} D^2(\vec{r})\, dV_i \quad (13)$$

Therefore, the use of Eq. (11) can lead to bias in the results.

To quantify this bias, $D_i$ and $\overline{D_i^2}$ were evaluated according to Eqs. (12) and (13), respectively, for a voxel containing an AuNP subject to an ionizing photon interaction. Following the approach presented in (Rabus and Thomas, 2025), the local dose is written as the sum of the dose without AuNP and the dose from a single AuNP undergoing an ionization. This results in the following expressions for the mean dose and the mean square of the dose in the voxel (Eqs. (14) to (16)):

$$D_i = D_w + \overline{D_k} \; ; \overline{D_k} = \frac{1}{V_v}\int_{r_p}^{\infty} D_1(r) w_k(r) dr \quad (14)$$

$$\overline{D_i^2} = D_i^2 + \overline{D_k^2} - \overline{D_k}^2 \quad (15)$$

$$\overline{D_k^2} = \frac{1}{V_v}\int_{r_p}^{\infty} D_1(r)^2 w_k(r) dr \qquad (16)$$

The lower integration limit $r_p$ is the radius of the spherical AuNP, $D_w$ is the dose without AuNPs, $D_1(r)$ is the additional dose resulting from the interaction in the AuNP at a radial distance $r$ from its center, and $w_k(r)$ is a geometric weighting function. The integral over $w_k(r)$ gives the volume $V_v$ of the voxel, and $w_k(r)dr$ corresponds to the probability of finding a point in the voxel at a certain radial distance from the AuNP. The index $k$ indicates that the weighting function depends on the position of the AuNP in relation to the voxel.

The weighting functions were determined by Monte Carlo sampling uniformly distributed points in the voxel and determining the frequency distribution of the radial distance to AuNP positions in the center, the centers of the faces and edges, and the corners of the voxel.

## 2.5 Dose around a 5 nm AuNP undergoing an ionization

The data for the dose around an AuNP were taken from the simulations of (Rabus and Thomas, 2025), using the results for energy imparted around a spherical AuNP with a diameter of 5 nm which was closest to the diameter of 4.89 nm used by (Antunes et al., 2025a). As described in (Thomas et al., 2024), the data were generated in a two-step simulation.

In the first step, the photon and electron fluence were determined in a scoring cylinder of 100 µm length in a water phantom with Geant4-DNA (Version 11.1.1) (Incerti et al., 2010a, 2010b, 2018; Bernal et al., 2015; Sakata et al., 2019). The center of the cylinder was 100 µm away from the phantom surface. The cylinder had a radius of 10 cm, and the surrounding cylindrical phantom had a height of 100 cm and a radius of 50 cm to ensure a saturated contribution of backscattered photons.

The incident radiation was a parallel photon beam with a radius of 10 cm (Thomas et al., 2024). The simulations were performed with Geant4-DNA option 2, using track structure mode in the scoring cylinder and within a concentric cylindrical region with outer surfaces at a distance of 50 µm from those of the cylinder. Electron transport was suppressed outside this region.

In the second simulation, the photon and electron fluence spectra from the first simulation were used for the source under the assumption of isotropic and laterally uniform irradiation. The electron transport was simulated in track structure mode in the AuNP and the surrounding water. In the simulations of (Rabus and Thomas, 2025), the histories of primary particles leaving the AuNP were terminated, so that only the energy imparted from secondary particles was scored in spherical shells with a thickness of 5 nm starting from the AuNP surface. The contributions resulting from photons and electrons impinging on the AuNP were added using the ratios of the particle fluences per primary fluence from the first simulations as weights.

## 2.6 Cell survival prediction for 100 kV x-rays

To estimate the overall effect of cell geometry and AuNP distribution in the cell on predicted survival rate, an approach analogous to that of Rabus and Thomas (2025) was used. In this approach, the mean dose and mean square of the dose in the nucleus are determined from the dose distribution around a single AuNP and a geometric weighting function by Eqs. (14) and (15):

$$\overline{D} = D_w + \bar{n}_m \int_{r_p}^{\infty} D_1(r) w_1(r) dr \qquad (17)$$

$$\overline{D^2} \approx \overline{D}^2 + \bar{n}_m \int_{r_p}^{\infty} [D_1(r)]^2 w_1(r) dr \qquad (18)$$

Where $\bar{n}_m$ is the number density of AuNPs undergoing ionizing interaction, which is proportional to $D_w$, and $w_1(r)$ is the weighting function. This weighting function can be interpreted as the probability of finding an AuNP at a given radial distance from a point in the nucleus, averaged over all points in the nucleus.

It should be noted that in Eq. (15) the contribution of AuNP pairs is approximated by the square of the mean dose. This is justified since the difference between the two in (Rabus and Thomas, 2025) was found to be small.

The analytical weighting functions presented in (Rabus and Thomas, 2025) refer to a spherical geometry and therefore cannot be applied to the cell models of (Antunes et al., 2025a). Therefore, the weighting functions for the whole cells were determined by random sampling pairs of points, one of which was uniformly distributed in the nucleus and the second uniformly distributed either in the whole cell or also in the nucleus.

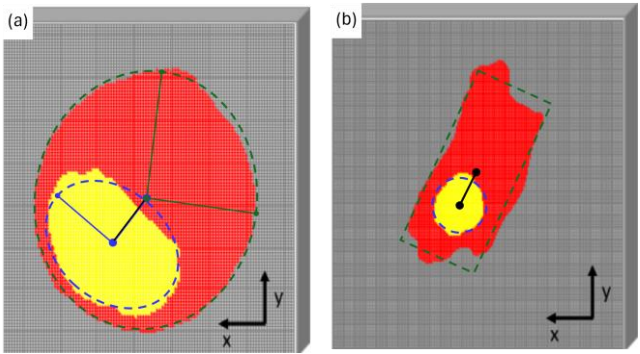

Fig. 1: Cross-sections of the models for (a) a cell in suspension and (b) an adherent cell. Reproduced from (Antunes et al., 2025a) (copyright Antunes et al.) under the CC BY 4.0 license (http://creativecommons.org/licenses/by/4.0/) with the following modifications: The two panels were cut out, rearranged, and the labels "(b)" and "(d)" were replaced by "(a)" and "(b)". In addition, dashed contours, their center points and lines indicating their half axes were added.

This resulted in two weighting functions $w_c$ and $w_n$, which were normalized so that their integral gave the volume of the cell. If a proportion $p$ of the number of AuNPs in the cell is located in the cell nucleus, the corresponding weighting function $w_p$ is obtained using Eq. (19).

$$w_{1p}(r) = \frac{V_c w_c - V_n w_n}{V_c - V_n} + p \frac{V_c(w_n - w_c)}{V_c - V_n} \quad (19)$$

Approximate geometries were defined based on the cross-sectional views shown in (Antunes et al., 2025a), reproduced here in Fig. 1. The green dashed ellipse was manually fitted to the cross-section of the cell in suspension in Fig. 1(a) and the green-dashed rectangle to that of the adherent cell in Fig. 1(b). Comparison of the areas of these shapes on paper with the cross-sectional areas given by (Antunes et al., 2025a) gave the length scales in the two diagrams. The heights of the shapes were then determined from the volumes of the cells, assuming the cells have the shape of an elliptical cylinder and a cuboid, respectively (Rabus and Mkanda, 2025).

For the floating cell, the blue dashed ellipse was fitted to the yellow area in Fig. 1(a), which represents the cell nucleus. The shape of an elliptical cylinder was also assumed for this nucleus, the height of which was calculated from the nucleus volume given in (Antunes et al., 2025a). For the adherent cell, the nucleus was assumed to be spherical. The blue dashed circle in Fig. 1(b) indicates the cross-section of the sphere approximating the nucleus. For simplicity, the center of the nucleus was assumed to be at half the height of the cell in both cases. The analysis in (Rabus and Mkanda, 2025) has shown that there is an offset between the centers of cells and nuclei. However, is was not the intention to exactly reproduce the cell models of (Antunes et al., 2025a), but to obtain an approximate and easy to implement geometry. The values used for the different geometry parameters are listed in Table 1.

## 3. Results

### 3.1 Photon fluence at the MWP

The photon spectrum shown in Fig. 2 consists of the two dominating $^{60}$Co gamma lines and photons over a broad energy range at significantly lower fluence, which has been multiplied by a factor of 50 to make it visible. The peak structure above 200 keV corresponds to the lowest energies of Compton-backscattered $^{60}$Co gamma photons, the peak at 511 keV is from annihilation of positrons (produced in pair-creation processes). The proportion of the two main gamma lines to the total spectrum integral amounts to 0.59 resulting in a value for $G_{sec}$ 0.69.

For evaluating the other quantities entering Eq. (2), the distance between the $^{60}$Co source and the MWP is estimated from information given in (Cardoso, 2023), according to which the $^{60}$Co sources are located 18 cm above the MWP and at 10 cm radial distance from its center. The resulting distance $d_a + d_s$ is about 21 cm. Using the area of the MWP of 57.63 cm² stated in (Antunes et al., 2025a), the fraction of the solid angle is thus obtained as about 0.01.

The linear attenuation coefficients $\mu_s$ and $\mu_a$ appearing in Eq. (2) were estimated from mass-attenuation coefficients retrieved from the NIST XCOM online database (Berger et al., 2010). They are listed in Supplementary Table 1 for the photon energies corresponding to the two $^{60}$Co gamma lines for cobalt and air. It was assumed that the $^{60}$Co source consists of cobalt only and the density of air was assumed to be that of dry air at sea level and standard conditions.

Assuming the average beam path of photons within the $^{60}$Co source to be between 0.5 cm and 1 cm, the values for $n_{pd}$ resulting from Eq. (2) are between 0.022 and 0.028. For comparison, if the source was a point source emitting only the two main gamma lines of $^{60}$Co, the correction factor would be 0.02. Therefore, the presence of the lower energy photons in the spectrum has only a minor influence on the estimated correction.

Table 1: Dimensions of the shapes used as approximations to the cell models of Antunes et al. (2025).

| parameter | suspension | adherent |
|---|---|---|
| long axis or side of cell / μm | 18.24 | 26.00 |
| short axis or side of cell / μm | 16.40 | 12.70 |
| cell thickness in z-direction / μm | 17.00 | 8.40 |
| long axis of nucleus / μm | 10.90 | 8.56 |
| short axis of nucleus / μm | 7.76 | 8.56 |
| nucleus thickness / μm | 9.50 | 8.56 |
| nucleus offset in x-y-plane / μm | 2.84 | 4.14 |
| rotation of nucleus / deg | 55 | - |

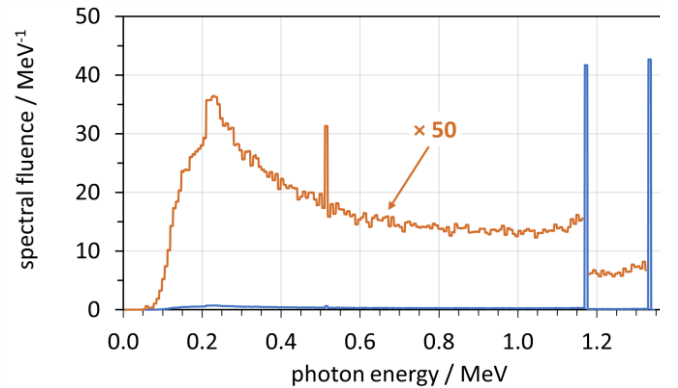

Fig. 2: Photon energy spectrum at the PTB $^{60}$Co calibration facility at 1 m distance from the source (10 cm×10 cm reference field). The data at energies different from the $^{60}$Co gamma lines have been multiplied by 50 for better visibility.

However, the determination of the correction factor $n_{pd}$ according to Eq. (4) gave significantly smaller values of $3.6\times10^{-3}$ for the photon spectrum in Fig. 2 and $2.9\times10^{-3}$ for the two $^{60}$Co gamma lines.

*3.2 Dose build-up*

Fig. 3 shows the dependence on the beam radius of the dose to fluence ratio resulting from irradiation of a water phantom with the photon spectrum shown in Fig. 2. The dose is determined in cylinders of 100 μm length in beam direction, with centers located at depths of 50 μm, 5 mm and 1 cm. The data shows saturation behavior with an increasing beam radius. The saturation values are indicated by the horizontal lines and differ by a factor of about 10 between the volume near the surface and those at depths of 5 mm and 1 cm. For the volume near the surface, the saturation value for large beam diameter is by about a factor of 5 higher than the value obtained with a beam of diameter of 10 μm. The saturation values at depths of 5 mm and 1 cm are thus by a factor of about 50 higher than what is obtained with a beam of diameter of 50 μm in the volume near the surface.

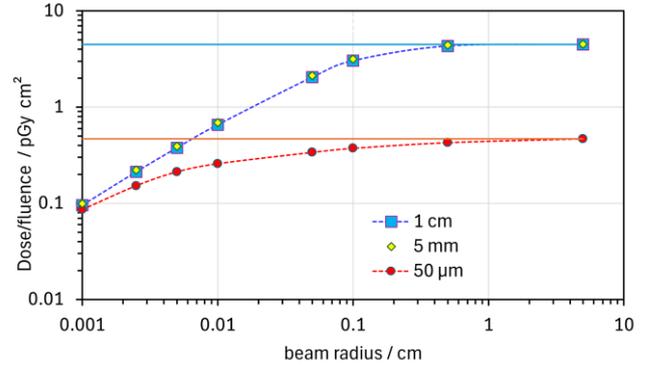

Fig. 3: Dependence of the dose per fluence on the radius of the beam for different depths in the phantom as given in the legend for the photon spectrum shown in Fig. 2. The dose values are averaged over 100 μm in depth. The horizontal lines indicate the saturation values for large beam radius.

*3.3 Photon and electron spectra at the cells*

The photon fluence spectra obtained in the PENELOPE simulations are shown in Fig. 4(a) where the label "primary" corresponds to the incident photon spectrum. The other labels refer to the depth of the cylinder center under the surface and the radius, respectively. The data at energies exceeding 1 MeV (vertical gray line) were divided by 40 to avoid the two gamma lines at 1.173 MeV and 1.333 MeV go out of scale in Fig. 4(a). It should be noted that the energy bin width was 7 keV for the incident photon spectrum. Since the intrinsic energy width of the gamma lines is much smaller, a much larger factor would have to be used for a finer energy binning.

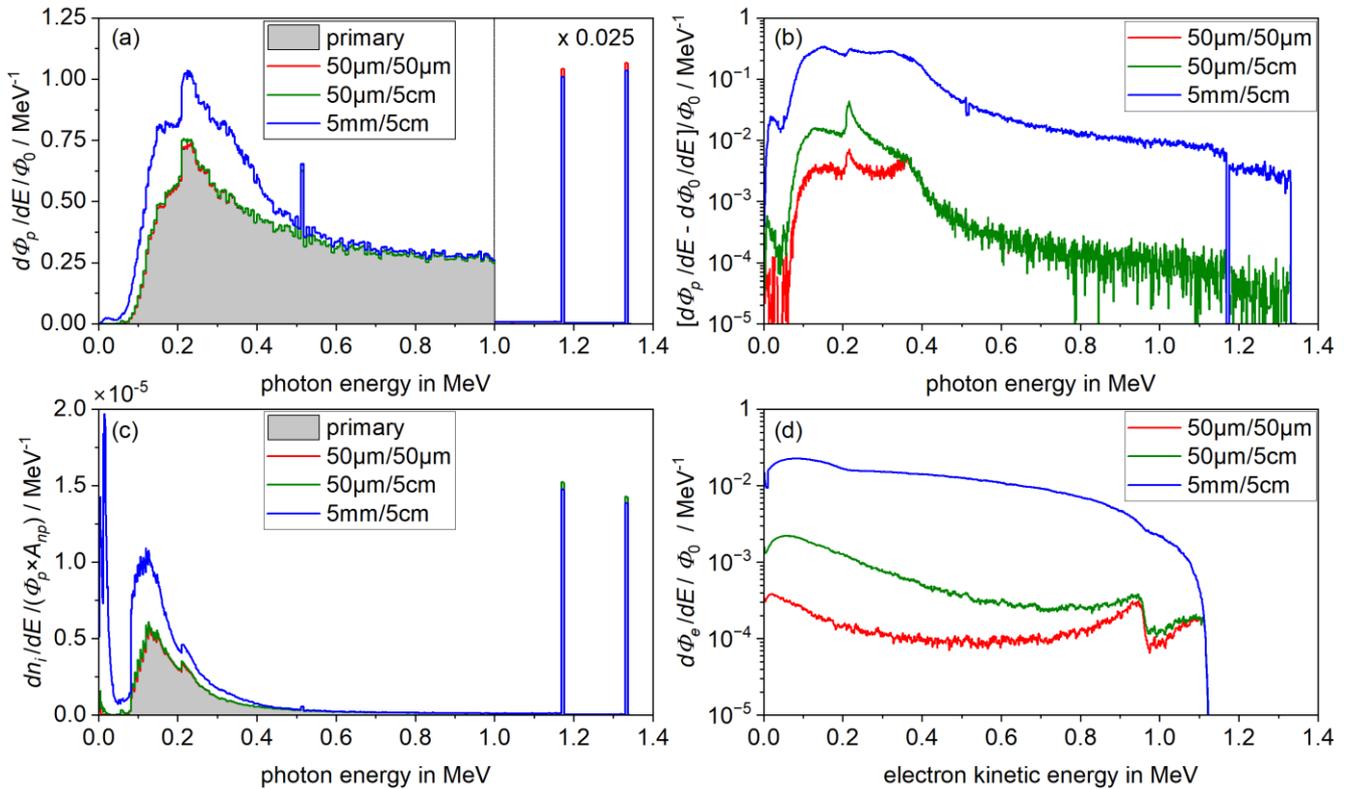

**Fig. 4**: (a) Fluence spectra of photons in a cylinder of radius 5 cm at 5 mm depth (blue), at the surface (black line), and in cylinders of radius 50 μm (red) and 5 cm (green) at 50 μm depth. The black, red, and green curves are indistinguishable. (b) Difference between the fluence spectra in the cylinders and the incident photon spectrum. (c) Number of ionizing interactions in the AuNP per photon for the fluence spectra shown in (a). (d) Electron fluence spectra in the three cylinders. All fluences are normalized to the primary photon fluence.

The data for cylinders of radius of 50 μm and 5 cm at 50 μm depth (red and green lines) are indistinguishable from the primary photon fluence (black line). For the cylinder of 5 cm radius at 5 mm depth (blue line), a clear increase in photon fluence can be seen at energies below 500 keV and a slight decrease at the two gamma lines.

Fig. 4(b) shows the differences between the fluence in the cylinders and the incident energy spectrum. Except for the two $^{60}$Co gamma lines at which the difference is negative, the photon fluence is slightly higher in the cylinders at 50 μm depth. For the cylinder of radius of 5 cm at a depth of 5 mm, the additional photon fluence at energies different from the those of the gamma lines is by one to two orders of magnitude higher than for the cylinders at 50 μm depth. However, the ratio of the total photon fluence to that of the $^{60}$Co gamma lines only increases from 1.69 for the incident spectrum to 1.90 for the cylinder of radius 5 cm at a depth of 5 mm.

The photon energy spectra shown in Fig. 4(a) were multiplied by the number of ionizations in the AuNP per fluence according to Eq. (5) and then normalized to the expected number of photons impinging on the AuNP. The results are shown in Fig. 4(c). Owing to the much smaller interaction cross-sections at higher energies, the contribution of the gamma lines per energy interval is now in the same order of magnitude as for low-energy photons. However, the area under the curves is clearly dominated by the part of the curve at photon energies below 500 keV.

Fig. 4(d) shows the electron fluence spectra for the three cylinders. For the cylinders at a depth of 50 μm the electron fluence at energies below 950 keV is increased for the larger cylinder of radius of 5 cm by up to almost a factor of 10. The electron fluence in the cylinder with radius of 5 cm at a depth of 5 mm is increased by about an order of magnitude at electron energies below 350 keV and up to a factor of about 50 at higher energies.

### 3.4 Number of interactions in the simulations

The ratios of the total fluences in the three cylinders to the primary photon fluence are shown in Table 2, which also contains results for a cylinder at a depth of 1 cm. The increase in photon fluence with depth and cylinder radius stays below 10 %. For the electron fluence, an increase of about a factor of 5 is obtained for the cylinders at 50 μm depth when the radius increases from 50 μm to 5 cm. A further increase of electron fluence by more than an order of magnitude is obtained when the cylinder is at depths of 5 mm and 1 cm. This increased electron fluence is the cause of the dose increase observed in Fig. 3.

Table 2 also shows the expected number of ionizations in an AuNP of diameter of 4.89 nm hit by a photon, the expected total number of events in which a photon or an electron hit one of the 4326 AuNPs in the cell, and the expected number of events in which an ionization by a photon occurs in an AuNP. The last three values were estimated using Eqs. (7) and (8) and correspond to a simulation with $10^7$ primary particles and the cell in suspension.

While the expected number of events (out of $10^7$) where a photon hits an AuNP is always in the order of 3000, the corresponding number for electrons is less than 1 when the cylinder is near the surface. With increasing beam radius and depth, the number of events in which an electron hits an AuNP increases up to values in the order of 40.

The expected number of events in which ionizing photon interactions occur in an AuNP also increases with the beam diameter and depth in the phantom. However, this number is always in the permille range.

Table 2: Results for the photon and electron fluences at the surface and in cylinders of different radius $r$ located at different depth $d$. The height of the cylinder is 100 μm. $\Phi_p/\Phi_0$ and $\Phi_e/\Phi_0$ are the integral photon and electron fluences normalized to the photon fluence at the surface. The quantity $n_{i,1}$ is the expected number of ionizing photon interactions in a spherical AuNP of diameter of 4.89 nm for a fluence of one photon per cross-section $A_{np}$ of the AuNP calculated with Eq. (8). The last three columns give the expected number events in a simulation in which a photons hits an AuNP ($N_p$), an electron hits an AuNP ($N_e$), and an ionizing interaction of a photon occurs in an AuNP of 4.89 nm diameter ($N_i$). The values apply to a simulation with 4326 AuNPs in the cell and $10^7$ primary particles starting from an area of cross-section equal to that of the cell. The relative statistical uncertainty of the fluence ratios is about $5\times10^{-4}$, and a similar relative uncertainty is also assumed for the other quantities.

| region | $\Phi_p/\Phi_0$ | $\Phi_e/\Phi_0$ | $n_{i,1}$ | $N_p$ | $N_e$ | $N_i$ |
|---|---|---|---|---|---|---|
| surface | 1 | 0 | $1.13\times10^{-6}$ | 3385 | 0 | $3.83\times10^{-3}$ |
| $d = 50$ μm, $r = 50$ μm | 1.001 | $1.63\times10^{-4}$ | $1.14\times10^{-6}$ | 3389 | 0.55 | $3.87\times10^{-3}$ |
| $d = 50$ μm, $r = 5$ cm | 1.004 | $7.20\times10^{-4}$ | $1.17\times10^{-6}$ | 3398 | 2.44 | $3.98\times10^{-3}$ |
| $d = 5$ mm, $r = 5$ cm | 1.086 | $1.12\times10^{-2}$ | $2.13\times10^{-6}$ | 3677 | 37.9 | $7.81\times10^{-3}$ |
| $d = 1$ cm, $r = 5$ cm | 1.095 | $1.08\times10^{-2}$ | $2.33\times10^{-6}$ | 3706 | 37.1 | $8.65\times10^{-3}$ |

## 3.5 Bias in the voxel-based LEM

The weighting functions for the radial dose distribution around an AuNP for the mean dose in a voxel are shown in Fig. 5(a) for the voxel corresponding to the suspended cell. The different curves correspond to the cases where the AuNP is located in the center of the voxel (black curve), the centers of the faces (red and green), the centers of the edges (cyan and violet), and in the edges (blue). Depending on the location of the AuNP, the range over which the weighting function differs from zero varies between slightly larger than 300 nm (half diagonal of the voxel) for the center position to slightly more than 600 nm (full diagonal) for the edge position. Since the voxel is very prolate (sides along $x$ and $y$ are 210 nm as compared to 600 nm along $z$), very different weighting functions are found for the AuNPs in the centers of the large and small faces (perpendicular to the $x$- axis or $y$-axis and the z-axis, respectively). The same applies to the centers of the edges.

Fig. 5(b) shows the dose around an AuNP (Supplementary Fig. 1) multiplied by the weighting functions shown in Fig. 5(a). Here it can be seen that the relevant contributions are found within the first few hundreds of nanometers and therefore correspond mainly to the energy imparted by electrons with energies in the range of the gold M-shell Auger-Meitner electrons (below 3500 eV). The contribution to the mean square of the dose is dominated by the radial distances of few tens of nanometers, i.e. the range of the gold N-shell Auger-Meitner electrons.

The resulting contribution of voxels containing a single AuNP undergoing an ionization can be seen in the values listed in Table 3. The data are the additional contribution from the AuNP to the mean dose, the square of the mean dose, and the mean square of the dose in the voxel. For this purpose, the values obtained according to Eqs. (14) and (16) were divided by the number of voxels ($1.59 \times 10^4$) and multiplied by the number of voxels intersecting the AuNP. (I.e. by 2 for the center of a voxel face, 4 for AuNPs on an edge, and 8 for a corner.)

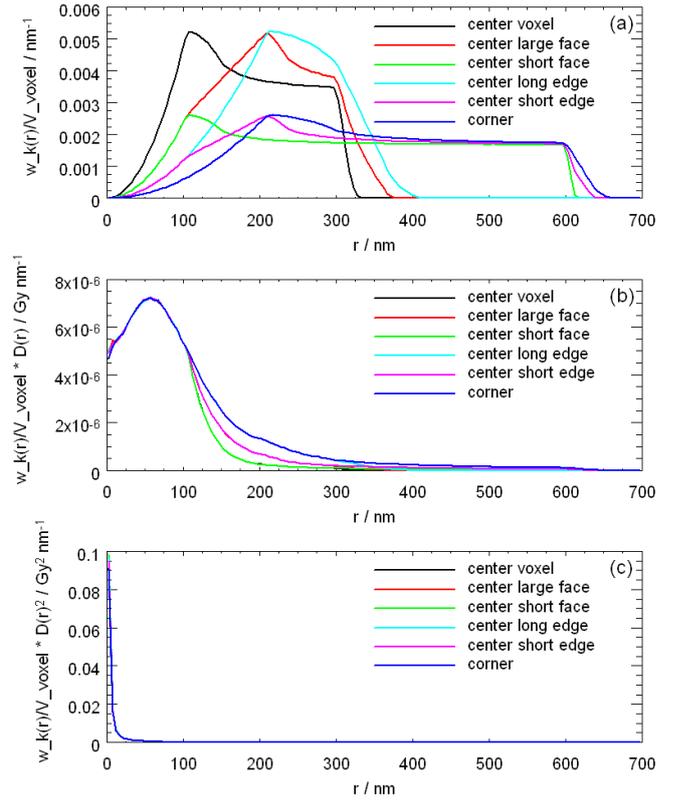

Fig. 5: (a) Weighting functions for the contribution of the dose at a radial distance from the AuNP to the mean dose and mean square of the dose in a voxel for different positions of the AuNP in the voxel (see legend). (b) contribution of different radial distance from the AuNP to the mean dose in the voxel. (c) contribution of different radial distances to the mean square of the dose in the voxel.

## 3.6 Estimated survival rate prediction for 100 kV x-rays

The weighting functions for the different proportions of AuNPs in the nucleus are shown in Supplementary Fig. 2(a) for the cell in suspension and in Supplementary Fig. 2(b) for the adherent cell. Owing to the smaller size of the nucleus of the adherent cell, the weighting function extends over a smaller range when all AuNPs are in the nucleus (violet curve) than for the cell in suspension. At the same time, due to the elongated shape of the adherent cell, larger radial distances may contribute to the dose if the AuNPs are in the cytoplasm.

Table 3: Effects of a single ionizing photon interaction in a GNP depending on the position of the GNP in the voxel: Mean dose $\overline{D}_k$ in the voxel; number $n_k$ of voxels containing the GNP; resulting contributions to the mean dose, the square of the mean dose and the mean square of the dose to the nucleus from these voxels, and contribution calculated in the approach of (Antunes et al., 2025a). The values apply to the model for the cell in suspension (nucleus volume of 420 µm³, voxel volume of $2.65 \times 10^{-2}$ µm³, number of voxels $n = 1.59 \times 10^4$).

| AuNP position | $\overline{D}_k$ / Gy | $n_k$ | $\frac{n_k}{n}\overline{D}_k$ / Gy | $\frac{n_k}{n}\overline{D}_k^2$ / Gy² | $\frac{n_k}{n}\overline{D_k^2}$ / Gy² |
|---|---|---|---|---|---|
| voxel center | 13.0 | 1 | $0.81 \times 10^{-3}$ | $1.05 \times 10^{-2}$ | 0.64 |
| large face center | 7.12 | 2 | $0.90 \times 10^{-3}$ | $0.64 \times 10^{-2}$ | 0.66 |
| small face center | 6.60 | 2 | $0.83 \times 10^{-3}$ | $0.55 \times 10^{-2}$ | 0.68 |
| long edge center | 4.01 | 4 | $1.01 \times 10^{-3}$ | $0.41 \times 10^{-2}$ | 0.67 |
| short edge center | 3.66 | 4 | $0.92 \times 10^{-3}$ | $0.34 \times 10^{-2}$ | 0.66 |
| corner | 2.10 | 8 | $1.06 \times 10^{-3}$ | $0.22 \times 10^{-2}$ | 0.64 |

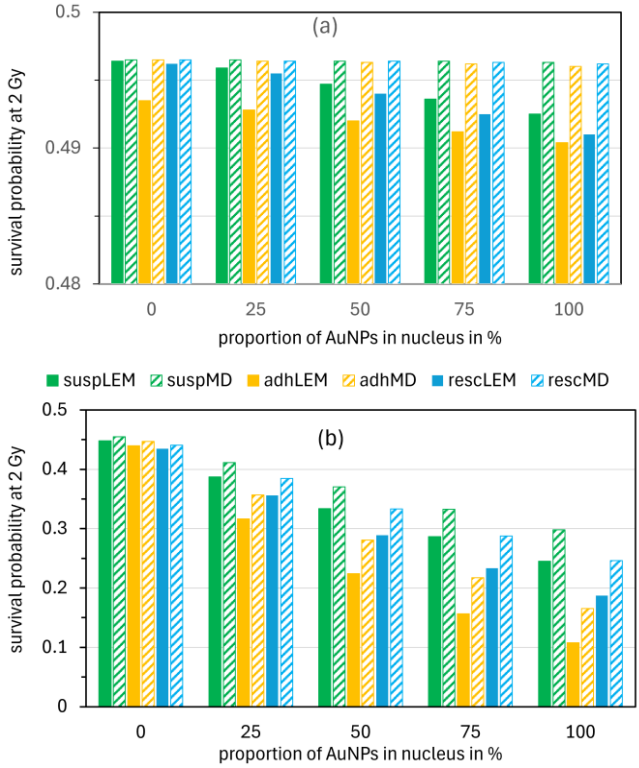

Fig. 6: Predicted survival rate at 2 Gy dose for the cell in suspension (green), the adherent cell (orange) and a rescaled cell in suspension, which has the same volume as the adherent cell (blue). The full columns are the predictions according to the LEM, the dashed bars indicate the survival rate predicted from the mean dose in the nucleus. (a) relates to 4326 AuNPs of 5 nm diameter and (b) to 4326 AuNPs of 50 nm diameter, in both cases irradiated by the mixed photon and electron field from a 100 kV incident x-ray spectrum at 100 μm depth in water.

The corresponding survival predictions for a dose of 2 Gy for cells with 4326 internalized AuNPs exposed to the mixed photon and electron field from a 100 kV primary x-ray beam at 100 μm depth in the phantom are shown in Fig. 6, where Fig. 6(a) relates to AuNPs of 5 nm diameter and Fig. 6(b) relates to AuNPs of 50 nm diameter. As in (Antunes et al., 2025a), the full columns correspond to the predictions according to the LEM and the dashed columns to the survival rate estimated from the mean dose in the nucleus. The different groups of columns correspond to different proportions of the number of AuNPs in the cell that are internalized in the nucleus.

The green columns refer to the cell in suspension and the orange columns refer to the adherent cell as shown in Fig. 1. The blue columns are for a rescaled version cell in suspension, which has the same volume as the adherent cell such as to assure the same mass fraction of gold in this cell and the adherent cell. It should be noted that the vertical scale in Fig. 6(a) shows only the range between 0.48 and 0.50 while the y-axis starts at 0 in Fig. 6(b).

## 4. Discussion

Simulation results for AuNPs can vary greatly depending on the parameters used (Vlastou et al., 2020), with the simulation geometry being particularly relevant for microscopic simulations of AuNPs (Zygmanski et al., 2013). (Antunes et al., 2025a) used a two-stage simulation approach in which a macroscopic and a microscopic simulation are linked via a phase space file (PSF) that is rescaled between the two simulation geometries. This approach has been used by several authors (Lin et al., 2014; Klapproth et al., 2021; Taheri et al., 2024; Antunes et al., 2025b), of which only a few considered charged particle equilibrium conditions (Antunes et al., 2025b; Taheri et al., 2024, 2025a). In (Antunes et al., 2025a) a simulation setup with a microscopic beam was used that was shown in (Rabus and Mkanda, 2025) to lead to biased results. These biased results entail biased conclusions regarding the dependence on cell shape and orientation of the cell survival prediction by the LEM.

### 4.1 Underestimation of LEM effects

Several other biases implied in the simulation setup and data analysis were quantified here. One of these is related to the use of the LEM with voxels as large as 0.2 μm. It can be seen from Table 3 that the contribution to the mean dose is negligible, and this also applies the contribution to the mean square dose estimated according to Eq. (11). However, the contribution to the mean square of the dose according to Eq. (16) is larger by a factor between 60 and 300 than that estimated from the mean dose in the voxel. It is worth noting that the values in the penultimate column of Table 3 are almost constant. This reflects the local nature of the LEM effect.

The contribution estimated from the approach of (Antunes et al., 2025a) varies by a factor of almost 5 for the different locations of the AuNP in the voxel. This is due to the fact that only the voxels containing the AuNP were considered here. The curves in Fig. 5(b) extend well beyond 100 nm, so that second closest neighbors and even more distant voxels also contribute to the mean dose, since their coordination number (second column of Table 3) is also higher. But even if the total contribution were as large as the sum of the values shown in the penultimate column of Table 3, the LEM-specific contribution to the mean square of the dose in the nucleus would still be underestimated by a factor of 20.

### 4.2 Bias in the primary photon fluence

By using Eq. (1) to calculate the number of photons hitting a cell, this number and thus the particle fluence was overestimated in the work of (Antunes et al., 2025a). The correction described by Eq. (2) takes into account the irradiation geometry and the scattering of the $^{60}$Co gamma photons on their way to the MWP. Since the actual spectrum obtained in the simulations was not reported in (Antunes et al., 2025a), the measured energy spectrum at a different irradiation facility was used as a proxy for the radiation field.

The resulting estimate for the expected number $n_{pd}$ of photons per decay of the $^{60}$Co source hitting the surface of the MWP was about 0.02 for a point source emitting only the $^{60}$Co gamma lines and about 0.025 for the photon spectrum shown in Fig. 2. The estimate based on Eqs. (3) and (4) was lower by a factor of about 6 to 7 in both cases. This indicates that the approximations underlying Eq. (2) are possibly too simplistic.

These approximations are based on the assumptions that the effect of the attenuation of the photons in the source was described by Beer's law with an average path in the source. The values used in the evaluation for this average path appear realistic but can be off by a factor of 2 or more. Likewise, an average distance was assumed that refers to the center of the MWP, while this distance is different for off-center positions on the MWP. It was also not taken into account that in the irradiation facility used by (Antunes et al., 2025a), the radiation is incident at an angle of about 30° off-normal, which leads to a reduction of the solid angle subtended by the MWP and results in a greater variation in the distances between the source and points on the surface of the MWP than at normal incidence.

In addition, the proportion of secondary photons in the radiation field may differ between the irradiation facility used by (Antunes et al., 2025a) and the one that generates the spectrum shown in Fig. 2. If the content of low energy photons were smaller or larger by a factor of 2, the estimated value for the correction factor $n_{pd}$ according to Eq. (2) would vary by about ±30 %.

Apart from the area of the MWP, the estimation of $n_{pd}$ according to Eqs. (3) and (4) does not require any details of the irradiation geometry as input. Therefore, it can be considered more realistic. It still depends on the photon spectrum, but the deviation between the case of a pure gamma emitter and the modulated spectrum in Fig. 2 is only 20 %. From this, it can be concluded that the correction factor $n_{pd} = 3.6 \times 10^{-3} \pm 20$ %.

This means that the number of photons that hit the cell at a certain dose was probably overestimated by a factor of about 270 in the work of (Antunes et al., 2025a). Conversely, this implies that the dose calculated from the simulations should have been smaller by a factor of 270, and thus smaller than the dose in the experiments. This paradox of a dose calculated from the number of photons, which in turn was derived from a dose, can be easily resolved if one recognizes that this discrepancy reflects the different irradiation conditions in the simulations compared to those in the experiments.

### 4.3 Dose buildup

(Antunes et al., 2025a) used a confined beam in the second stage of their simulations and placed the source near the cells. This corresponds to the case where the cells are near the surface. In the experiments used as a reference, one would expect most of the cells to be in the bulk of the volume of water containing them. As shown in Fig. 3, the dose in a cylinder of radius 10 μm at a depth of 50 μm was about 5 times higher when the cylinder had a radius of 5 cm (which corresponds to the size of the MWP) than in a cylinder with a radius of 10 μm. A cylinder at a depth of 5 mm with a radius of 5 cm receives a dose that is higher by a factor of about 50 than the dose in a cylinder with a radius of 10 μm near the surface.

The PENELOPE simulations were performed with a pencil beam and therefore do not correspond to the irradiation conditions in the experiment. However, they capture the essential features of the irradiation conditions, namely the production of secondary particles, of which photons with lower energies in particular have a higher probability of interacting with the AuNPs. Furthermore, the statistical expectation for the dose and fluence in a cylindrical volume around a pencil beam is the same as the expectation for these quantities in the center of a cylindrical beam of the same cross-section. This means that the dose per fluence values in Fig. 3 are the same as those that would be obtained in the center of a beam of the radius indicated on the *x*-axis.

The variation of the radial dose profile is negligible across dimensions of 10 μm for a beam diameter in the mm range. Therefore, it can be assumed that the dose estimates for large beam diameters, indicated by the horizontal lines in Fig. 3, are correct. In contrast, the values at the smallest beam radius are expected to be overestimated, as there is a strong radial fall-off. This mean that the difference between the dose from the simulation of a cell irradiated near the surface with a beam of 10 μm radius and the dose experienced by a cell in an extended beam at greater depths may even be larger than the value of 50 estimated here.

### 4.4 Radiation field at the cells

Fig. 4(a) shows that the photon fluence spectra in the cylinders near the surface (red and green lines) are almost indistinguishable from the primary fluence spectrum. The minor deviations only become visible when the differences are considered as shown in Fig. 4(b). Fig. 4(b) shows that a larger cylinder or beam radius results in an increased photon fluence particularly in the energy range between 100 keV and 400 keV due to backscattered Compton photons. The photon fluence in 5 mm depth shows a significant enhancement in the whole energy range except for the two gamma lines. The enhancement is due to both forward and backscattered Compton photons; the interactions leading to the generation of forward scattered photons result in the observed decrease in fluence at the gamma lines.

Fig. 4(c) shows that there are more interactions in AuNPs with cells deeper in the volume. This is related to the increase of photons at energies below 400 keV and to the energy dependence of the cross-section for photoelectric absorption, which decreases with increasing photon energy according to a power law with an exponent between -3 and -4.

The fluence of the secondary electrons shown in Fig. 4(d) shows an increase by orders of magnitude between the surface and the bulk. For the cylinders near the surface, a larger photon beam size leads to electrons generated by backscattered photons and a resulting increase in electron fluence compared to a beam of 10 μm, where the fluence of lower energy electron is preferentially increased. This increase is in the order of about a factor of 5 and reflects the increase in dose by a factor of 5 between 10 μm beam size and 5 cm beam size.

For the cylinder at 5 mm depth, the increase in fluence extends over the complete energy range, which is due to the contribution from electrons scattered in forward directions. These have larger energies because the photons producing them are scattered backwards, which maximizes the momentum and energy transfer. Again, the order of magnitude difference between the green and the blue electron fluence curves in Fig. 4(d) explains the factor of about 10 between the saturation values of dose per fluence in Fig. 3, since the energy deposits that make up the absorbed dose are mainly from electron interactions.

As can be seen from Table 2, the total photon fluence increases only slightly with an increasing beam radius and increasing depth. On the contrary, the total fluence of electrons increases by a factor of about 4.5 for the cylinder of radius 5 cm near the surface compared to the corresponding cylinder of radius 50 μm. The same cylinder placed at 5 mm depth receives an electron fluence by a factor of about 15 higher. At 1 cm depth, the electron fluence decreases again. This is because transient electron equilibrium prevails at this depth, and the electron fluence therefore decreases following the trend of the photon beam attenuation.

The expected number of events in which a photon or an electron hits an AuNP are proportional to the corresponding fluences. While for all depths and beam radii the number of events with a photon hitting an AuNP is around 3000 for the considered number of 4326 AuNPs of 4.89 nm diameter in the cell and $10^7$ primary particles simulated, the number of events with electron hits remains two orders of magnitude smaller, even though it increases by a factor of about 75 with increasing depth. The number of events with ionizing interaction of a photon in an AuNP more than doubles between 50 μm depth and 5 mm depth. However, the values at 5 mm and 1 cm depths are still below 0.008. This means that an event with an ionizing photon interaction is only expected once per 125 simulations. For cells near the surface, such an event is only expected once per 250 simulations.

It is worth noting that when irradiated with an extended beam, the dose per fluence increases by an order of magnitude between the cylinders at the surface and at 5 mm depth (Fig. 3), while the expected number of ionizing photon interacions in an AuNP only doubles (Table 2). This mean that the number of interactions per dose for the cylinder near the surface is higher by a factor of about 5 than for the cylinder at 5 mm depth.

This implies a caveat for all results from simulation studies in which a radiation source was placed near the AuNP, namely that they are not representative of the situation in experiments. In experiments with cells in suspension, only a small percentage are close to the surface and the experiment averages over all cells. When cells are adherent and are irradiated with only a small layer of medium, the cells are closer to the surface and therefore their surviving fraction will reflect the higher number of interactions occurring in AuNPs per dose compared to cells in suspension.

However, great care must be taken when specifying the dose and to which a certain survival level refers. The dose applied in the experiments corresponds to the dose determined by a measurement with a dosemeter in the irradiation facility only if similar irradiation conditions are ensured, e.g. by using a bolus as a dose build-up layer. Otherwise, one compares apples with oranges.

Similarly, the simulations should always ensure charged particle equilibrium (or include a corresponding correction) and a realistic radiation field, as in (Antunes et al., 2025b; Rabus et al., 2021; Taheri et al., 2025a, 2025b; Thomas et al., 2024). For the experiments considered by (Antunes et al., 2025a), this implies the average dose and the particle fluence in the MWP must be determined in the first simulation.

*4.5 Dependence of predicted survival on cell shape.*

It is worth noting that the number of ionizing interactions per dose resulting from (Table 2) is about $1.0 \times 10^{-7}$ Gy$^{-1}$ for AuNPs at 5 mm depth. This value is more than two order of magnitude smaller than that for the 100 kV x-ray spectrum of (Thomas et al., 2024), which is $3.2 \times 10^{-5}$ Gy$^{-1}$. For the same number of AuNPs of 5 nm diameter per cell as assumed by (Antunes et al., 2025a) the survival rates at a dose of 2 Gy in Fig. 6(a) predicted from the mean dose in the nucleus are almost the same for all three cell geometries and all proportions of AuNPs in the nucleus. In all cases, a small decrease in survival probability is seen when the prediction is made by the LEM. As expected, the decrease in predicted survival increases with increasing proportion of AuNPs in the nucleus. However, the reduction in survival is only in the order of 1 %. Similarly, the differences between the adherent cell and the cell in suspension are only a few tenths of a percent. Reducing the volume of the cell in suspension to that of the adhenrent cell reduces the differences. However, the uncertainties of the predictions are also in the order of 1 % so that the differences seen in Fig. 6(a) are not statistically significant.

For the case of 4326 AuNPs with a diameter of 50 nm shown in Fig. 6(b), a reduction in the survival rate by 20 % can already be seen if the AuNPs are only in the cytoplasm. This greater reduction compared to the case of 5 nm AuNPs reflects the increased mass fraction of gold in the cell (higher by a factor of 1000). An increasing proportion of the AuNPs in the nucleus leads to a decreasing survival rate as shown by the decreasing height of the dashed columns. This indicates a corresponding increase in the dose in the nucleus. However, the decrease in the predicted survival fraction by using the LEM is much more pronounced. (Note that the $y$-axis in Fig. 6(b) starts at 0.) For the case of irradiation with 100 kV $x$-rays and a mass fraction of gold corresponding to 4326 AuNPs of 50 nm diameter in the cells, there is a clear effect of cell geometry, i.e. size and shape of the cell and relative position of the nucleus. The comparison of the orange and blue columns in Fig. 6(b) shows that this is the case even when the adherent cell and the cell in suspension have the same volume.

However, it should be noted that this statement is based on only one sample from each population of adherent cells and cells in suspension General conclusions about the extent of the influence of cell shape on the effects of AuNPs under irradiation require examination of representative samples from both cell populations to rule out the possibility that the results are coincidential.

In any case it can be stated that for the concentration of AuNPs of 5 nm diameter considered by (Antunes et al., 2025a), no significant effects are found for 100 kV x-rays. Since the interaction probabilities for $^{60}$Co irradiation are much smaller, these findings confirm that the effects observed in the experiments are not consistent with dosimetric effects of AuNPs. It cannot be judged whether this is due to an incorrect determination of the amount of internalized gold in the cells or whether this indicates that the effects indicate that there is radiosenitization due to the AuNPs that has nothing to do with their radiation absorption properties.

*4.6 AuNP interactions*

In (Rabus and Mkanda, 2025) it was estimated that (Antunes et al., 2025a) used a factor of about 30 times more photons than the $10^7$ in the simulations to estimate the fluence corresponding to a dose of 2 Gy. The results presented above indicate that this number was overestimated by a factor of about 270. So instead of multiplying by 30, the number of photons should have been divided by about 9. This means that the values given in the last three columns of Table 2 for a dose of 2 Gy should be by almost an order of magnitude smaller. Therefore, at a dose of 2 Gy the number of events in which an electron hits an AuNP should be about 4, and the number of events with ionizing photon interactions in an AuNP should be below 0.001 for cells at 5 mm depth.

This estimate applies to both simulations and experiments. Thus, if the cells in the experiments used as a reference by (Antunes et al., 2025a) contained 4326 AuNPs with a diameter of 4.89 nm, at a dose of 2 Gy, photon interaction with an AuNP occurred only in about one in a thousand cells. This means that, from a dosimetric point of view, there should be no discernible effect of the AuNPs on cell survival.

However, the experiments showed a notable effect of AuNPs on cell survival, which decreased from 46.6 % to only 2.7 % for PC3 cells at a dose of 2 Gy (Marques et al., 2022). We can only speculate about possible explanations for this contradiction.

As pointed out in (Rabus and Mkanda, 2025), the unexpectedly high number of electrons emitted from AuNPs and the occurrence of photon interactions in the simulations could be explained if the AuNP diameter was a factor of 10 larger than the assumed 4.89 nm. However, the TEM images in (Marques et al., 2022) confirm this average diameter of the AuNPs.

Another possibility would be that an error occurred when the gold content was determined. Most of the results reported in (Marques et al., 2022) are for a gold concentration of 36 µg/ml, which corresponds to a mass fraction of gold of only $3.6 \times 10^{-5}$. (Marques et al., 2022) also reported that the gold concentration in the PC3 cells was $4.81 \times 10^{-6}$ ng per cell, which corresponds to the value of 4326 AuNPs with a diameter of 4.79 nm. However, the cell in suspension has a volume of $4.0 \times 10^{-9}$ cm³. Therefore the gold concentration in this cell was about 1.2 µg/ml, which is about a factor of 30 lower than in the solution. As this figure does not indicate increased uptake by the cells, it could be that the gold content was higher than the stated value.

Yet another possibility is that the AuNPs act as radiosensitzer owing to their chemical or biochemical characteristics and that this effect is not related to their photoabsorption properties. The fact that (Marques et al., 2022) found the same cell viability for irradiated PC3 cells with AuNP concentrations of 36 µg/ml and 75 µg/ml supports this explanation. In this case all attempts to understand the reduction in cell survival based on radiophysical simulations are in vain.

## 5. Conclusions

The aim of this work was to estimate the bias introduced by the simulation setup and scoring approach used in (Antunes et al., 2025a). Determining the LEM-specific contribution to the mean square of the dose in the nucleus from the mean dose values per voxel underestimates this contribution by one to two orders of magnitude for the voxel size related to the suspended cell. In the case of $^{60}$Co irradiation, the use of a beam confined to cell dimensions and emitted from a source close to the cell leads to an underestimation of the dose by a factor of about 5. Compared to cells irradiated at depths in the phantom where charged particle equilibrium prevails, the underestimation is a factor of about 50 for the dose and a factor of about 2 for the probability of an ionizing photon interaction in an AuNP. Neglecting the solid angle of the multi-well plate leads to an overestimation of the number of photons hitting the cells of about 300. These estimates corroborate the concerns pointed out in the first paper regarding the reliability of the evidence reported by (Antunes et al., 2025a), which were based only on inconsistencies in the reported data.

More generally, the present findings show that simulations must ensure a realistic approximation of the radiation field and the dose distribution in the macroscopic sample in order to achieve results comparable with experiments. A strong bias in the simulation results can be avoided by scoring the particle fluence in a volume and not on a surface. This also allows the determination of the dose without AuNPs under conditions of charged particle equilibrium. It is then possible to assess the additional contribution to radiation effects due to the presence of AuNPs. In any case, plausibility checks of the results should always be performed on the basis of existing knowledge to avoid reporting unrealistic results.

Another general conclusion is that the use of the LEM with voxel averaged dose values can easily lead to unrealistic estimates. However, in a straightforward further development of the approach used here, a voxelized cell geometry can be used to obtain less approximate estimates of the expected contributions of AuNPs to cell survival. The approximate results obtained here show the continuous trends that would be expected for the variation of survival with concentration of AuNPs in the cell nucleus. They also indicate that for 5 nm AuNPs at the concentrations considered by (Antunes et al., 2025a) no significant change in survival is expected under 100 kV x-ray irradiation, and this also applies to $^{60}$Co irradiation due to the lower photon interaction probability. Significant effects of different survival predictions for the two cell geometries were found for the same number density of 50 nm AuNPs, i.e. a factor of 1000 higher gold content, even if the cells had the same volume. However, these are results for one cell of each type. Sustainable conclusions require that this is investigated for statistically relevant samples from both populations.

## 6. References

Antunes, J., Pinheiro, T., Marques, I., Pires, S., Botelho, M.F., Sampaio, J.M., Belchior, A., 2025a. Do cell culturing influence the radiosensitizing effect of gold nanoparticles: a Monte Carlo study. EJNMMI Phys. 12, 41. https://doi.org/10.1186/s40658-025-00746-3

Antunes, J., Pinto, C.I.G., Campello, M.P.C., Santos, P., Mendes, F., Paulo, A., Sampaio, J.M., 2024. Utility of realistic microscopy-based cell models in simulation studies of nanoparticle-enhanced photon radiotherapy. Biomed. Phys. Eng. Express 10, 025015. https://doi.org/10.1088/2057-1976/ad2020

Antunes, J., Rabus, H., Paulo, A., Sampaio, J.M., 2025b. Chemical mechanism in gold nanoparticles radiosensitization: A Monte Carlo simulation study. Radiat. Phys. Chem. 232, 112637. https://doi.org/10.1016/j.radphyschem.2025.112637

Berger, M.J., Hubbell, J.H., Seltzer, S.M., Chang, J., Coursey, J.S., Sukumar, D.S., R. and Zucker, Olsen, K., 2010. XCOM: Photon Cross Section Database version 1.5) Available at: http://physics.nist.gov/xcom (Gaithersburg, MD: National Institute of Standards and Technology) [WWW Document]. https://doi.org/10.18434/T48G6X

Bernal, M.A., Bordage, M.C., Brown, J.M.C., Davídková, M., Delage, E., El Bitar, Z., Enger, S.A., Francis, Z., Guatelli, S., Ivanchenko, V.N., Karamitros, M., Kyriakou, I., Maigne, L., Meylan, S., Murakami, K., Okada, S., Payno, H., Perrot, Y., Petrovic, I., Pham, Q.T., Ristic-Fira, A., Sasaki, T., Štěpán, V., Tran, H.N., Villagrasa, C., Incerti, S., 2015. Track structure modeling in liquid water: A review of the Geant4-DNA very low energy extension of the Geant4 Monte Carlo simulation toolkit. Phys. Med. 31, 861–874. https://doi.org/10.1016/j.ejmp.2015.10.087

Cardoso, S., 2023. Radiosensitizers for Cancer Radiation Therapy (Master Thesis). Técnico Lisboa, Lisboa, Portugal.

Incerti, S., Baldacchino, G., Bernal, M., Capra, R., Champion, C., Francis, Z., Guatelli, S., Guèye, P., Mantero, A., Mascialino, B., Moretto, P., Nieminen, P., Rosenfeld, A., Villagrasa, C., Zacharatou, C., 2010a. The Geant4-DNA project. Int. J. Model. Simul. Sci. Comput. 1, 157–178. https://doi.org/10.1142/S1793962310000122

Incerti, S., Ivanchenko, A., Karamitros, M., Mantero, A., Moretto, P., Tran, H.N., Mascialino, B., Champion, C., Ivanchenko, V.N., Bernal, M.A., Francis, Z., Villagrasa, C., Baldacchino, G., Guèye, P., Capra, R., Nieminen, P., Zacharatou, C., 2010b. Comparison of GEANT4 very low energy cross section models with experimental data in water. Med. Phys. 37, 4692–4708. https://doi.org/10.1118/1.3476457

Incerti, S., Kyriakou, I., Bernal, M.A., Bordage, M.C., Francis, Z., Guatelli, S., Ivanchenko, V., Karamitros, M., Lampe, N., Lee, S.B., Meylan, S., Min, C.H., Shin, W.G., Nieminen, P., Sakata, D., Tang, N., Villagrasa, C., Tran, H.N., Brown, J.M.C., 2018. Geant4-DNA example applications for track structure simulations in liquid water: A report from the Geant4-DNA Project. Med. Phys. 45, e722–e739. https://doi.org/10.1002/mp.13048

Klapproth, A.P., Schuemann, J., Stangl, S., Xie, T., Li, W.B., Multhoff, G., 2021. Multi-scale Monte Carlo simulations of gold nanoparticle-induced DNA damages for kilovoltage X-ray irradiation in a xenograft mouse model using TOPAS-nBio. Cancer Nanotechnol. 12, 27. https://doi.org/10.1186/s12645-021-00099-3

Lin, Y., McMahon, S.J., Paganetti, H., Schuemann, J., 2015. Biological modeling of gold nanoparticle enhanced radiotherapy for proton therapy. Phys. Med. Biol. 60, 4149–4168. https://doi.org/10.1088/0031-9155/60/10/4149

Lin, Y., McMahon, S.J., Scarpelli, M., Paganetti, H., Schuemann, J., 2014. Comparing gold nano-particle enhanced radiotherapy with protons, megavoltage photons and kilovoltage photons: a Monte Carlo simulation. Phys. Med. Biol. 59, 7675–7689. https://doi.org/10.1088/0031-9155/59/24/7675

Marques, A., Belchior, A., Silva, F., Marques, F., Campello, M.P.C., Pinheiro, T., Santos, P., Santos, L., Matos, A.P.A., Paulo, A., 2022. Dose Rate Effects on the Selective Radiosensitization of Prostate Cells by GRPR-Targeted Gold Nanoparticles. Int. J. Mol. Sci. 23, 5279. https://doi.org/10.3390/ijms23095279

McMahon, S.J., Hyland, W., Muir, M., Coulter, J., Jain, S., Butterworth, K., Schettino, G., Dickson, G., Hounsell, A., O'Sullivan, J., Prise, K., Hirst, D., Currell, F., 2011a. Biological consequences of nanoscale energy deposition near irradiated heavy atom nanoparticles. Nat. Sci. Rep. 1, 18. https://doi.org/10.1038/srep00018

McMahon, S.J., Hyland, W.B., Muir, M.F., Coulter, J.A., Jain, S., Butterworth, K.T., Schettino, G., Dickson, G.R., Hounsell, A.R., O'Sullivan, J.M., Prise, K.M., Hirst, D.G., Currell, F.J., 2011b. Nanodosimetric effects of gold nanoparticles in megavoltage radiation therapy. Radiother. Oncol. 100, 412–416. https://doi.org/10.1016/j.radonc.2011.08.026

Melo-Bernal, W., Chernov, G., Barboza-Flores, M., Chernov, V., 2021. Quantification of the radiosensitization effect of high-Z nanoparticles on photon irradiated cells: combining Monte Carlo simulations and an analytical approach to the local effect model. Phys. Med. Biol. 66, 135007. https://doi.org/10.1088/1361-6560/abfce4

Rabus, H., 2024. Comment on "Reproducibility study of Monte Carlo simulations for nanoparticle dose enhancement and biological modeling of cell survival curves" by Velten et al. [Biomed Phys Eng Express 2023;9:045004]. Biomed. Phys. Eng. Express 10, 028002. https://doi.org/10.1088/2057-1976/ad1731

Rabus, H., Gargioni, E., Li, W., Nettelbeck, H., C., Villagrasa, 2019. Determining dose enhancement factors of high-Z nanoparticles from simulations where lateral secondary particle disequilibrium exists. Phys. Med. Biol. 64, 155016 (26 pp.). https://doi.org/10.1088/1361-6560/ab31d4

Rabus, H., Li, W.B., Villagrasa, C., Schuemann, J., Hepperle, P.A., de la Fuente Rosales, L., Beuve, M., Maria, S.D., Klapproth, A.P., Li, C.Y., Poignant, F., Rudek, B., Nettelbeck, H., 2021. Intercomparison of Monte Carlo calculated dose enhancement ratios for gold nanoparticles irradiated by X-rays: Assessing the uncertainty and correct methodology for extended beams. Phys. Med. 84, 241–253. https://doi.org/10.1016/j.ejmp.2021.03.005

Rabus, H., Mkanda, O.M., 2025. Do cell culturing influence the radiosensitizing effect of gold nanoparticles part 1: scrutinizing recent evidence for data consistency. ArXiv250619372 Physicsmed-Ph. https://doi.org/10.48550/ARXIV.2506.19372

Rabus, H., Thomas, L., 2025. Impact of metal nanoparticles on cell survival predicted by the local effect model for cells in suspension and tissue. Part 1: Theoretical framework. ArXiv250501909 Physicsmed-Ph 2505.01909 [physics.med-ph]. https://doi.org/10.48550/arXiv.2505.01909

Sakata, D., Kyriakou, I., Tran, H.N., Bordage, M.-C., Rosenfeld, A., Ivanchenko, V., Incerti, S., Emfietzoglou, D., Guatelli, S., 2019. Electron track structure simulations in a gold nanoparticle using Geant4-DNA. Phys. Med. 63, 98–104. https://doi.org/10.1016/j.ejmp.2019.05.023

Salvat, F., 2019. NEA/MBDAV/R(2019)1: PENELOPE-2018: A Code System for Monte Carlo Simulation of Electron and Photon Transport. Nuclear Energy Agency (NEA) of the Organisation for Economic Co-operation and Development (OECD), Paris.

Sung, W., Ye, S.-J., McNamara, A.L., McMahon, S.J., Hainfeld, J., Shin, J., Smilowitz, H.M., Paganetti, H., Schuemann, J., 2017. Dependence of gold nanoparticle radiosensitization on cell geometry. Nanoscale 9, 5843–5853. https://doi.org/10.1039/C7NR01024A

Taheri, A., Khandaker, M.U., Moradi, F., Bradley, D.A., 2024. A simulation study on the radiosensitization properties of gold nanorods. Phys. Med. Biol. 69, 045029. https://doi.org/10.1088/1361-6560/ad2380

Taheri, A., Khandaker, M.U., Rabus, H., Moradi, F., Bradley, D.A., 2025a. The influence of atomic number on the radiosensitization efficiency of metallic nanorods: A Monte Carlo simulation study. Radiat. Phys. Chem. 230, 112589. https://doi.org/10.1016/j.radphyschem.2025.112589

Taheri, A., Khandaker, M.U., Rabus, H., Moradi, F., Bradley, D.A., 2025b. The role of coating layers in gold nanorods' radioenhancement: A Monte Carlo analysis. Nanoscale Adv. 7, 3293–3307. https://doi.org/10.1039/D5NA00220F

Thomas, L., Schwarze, M., Rabus, H., 2024. Radial dependence of ionization clustering around a gold nanoparticle irradiated by x-rays under charged particle equilibrium. Phys. Med. Biol. 69, 185014. https://doi.org/10.1088/1361-6560/ad6e4f

Velten, C., Tomé, W.A., 2024. Reply to comment on 'Reproducibility study of Monte Carlo simulations for nanoparticle dose enhancement and biological modeling of cell survival curves' [Biomed Phys Eng Express 2023;9:045004]. Biomed. Phys. Eng. Express 10, 028001. https://doi.org/10.1088/2057-1976/ad17a9

Velten, C., Tomé, W.A., 2023. Reproducibility study of Monte Carlo simulations for nanoparticle dose enhancement and biological modeling of cell survival curves. Biomed. Phys. Eng. Express 9, 045004. https://doi.org/10.1088/2057-1976/acd1f1

Vlastou, E., Diamantopoulos, S., Efstathopoulos, E.P., 2020. Monte Carlo studies in Gold Nanoparticles enhanced radiotherapy: The impact of modelled parameters in dose enhancement. Phys. Med. 80, 57–64. https://doi.org/10.1016/j.ejmp.2020.09.022

Zygmanski, P., Liu, B., Tsiamas, P., Cifter, F., Petersheim, M., Hesser, J., Sajo, E., 2013. Dependence of Monte Carlo microdosimetric computations on the simulation geometry of gold nanoparticles. Phys. Med. Biol. 58, 7961–7977. https://doi.org/10.1088/0031-9155/58/22/7961

## Supplementary Figures and Tables

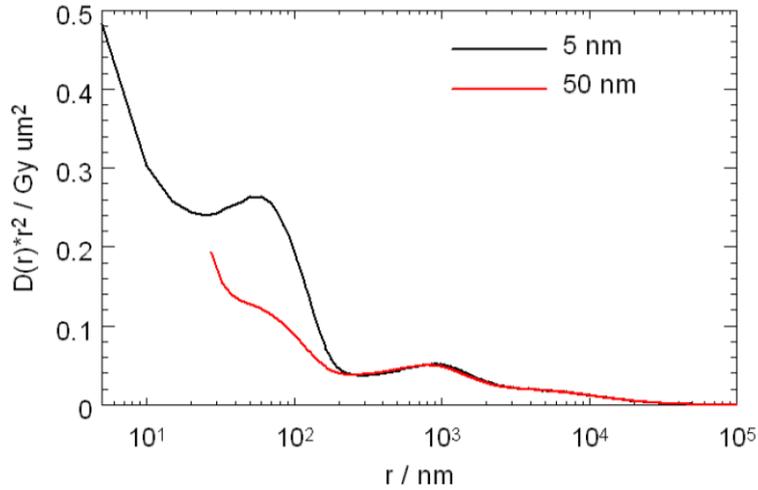

Supplementary Fig. 1: Dose around an AuNP of diameter 5 nm subject to an ionizing interaction in the mixed photon and electron field at 100 μm depths in a water phantom with an incident 100 kV x-ray beam (Rabus and Thomas, 2025).

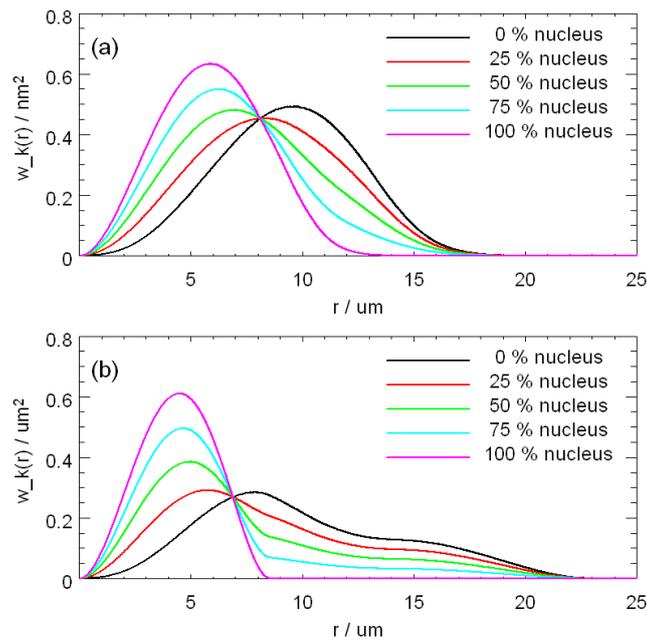

Supplementary Fig. 2: Radial weighting function for the extra dose around a GNP undergoing an ionizing photon interaction for (a) the cell in suspension and (b) the adherent cell. The different lines correspond to different proportions of the GNPs in the nucleus from the GNPs internalized in the cell.

Supplementary Table 1: Total mass-attenuations coefficients $\mu/\rho$ of cobalt and air at the energies of the two main $^{60}$Co gamma photons (Berger et al., 2010), and the resulting linear attenuation cofficients $\mu_s$ and $\mu_a$ of the source and the air in the irradiator, respectively. The $^{60}$Co source is assumed to consist pure Co (density 8.9 g/cm³) and the air to have a density of 1.2×10$^{-3}$ g/cm³.

| $E$ / MeV | $\left(\frac{\mu}{\rho}\right)_{Co}$ / cm²/g | $\left(\frac{\mu}{\rho}\right)_{air}$ / cm²/g | $\mu_s$ / cm$^{-1}$ | $\mu_a$ / cm$^{-1}$ |
|---|---|---|---|---|
| 1.173 | 5.22×10$^{-2}$ | 5.87×10$^{-2}$ | 0.484 | 7.04×10$^{-5}$ |
| 1.333 | 5.10×10$^{-2}$ | 5.49×10$^{-2}$ | 0.454 | 6.59×10$^{-5}$ |